\documentclass[3p,times,twocolumn]{elsarticle}

\usepackage{ecrc}

\volume{00}

\firstpage{1}

\journalname{Nuclear Physics B Proceedings Supplement}

\runauth{B. Lucini}

\jid{nuphbp}

\jnltitlelogo{Nuclear Physics B Proceedings Supplement}

\usepackage{amssymb}

\usepackage[figuresright]{rotating}

\begin{document}

\begin{frontmatter}



\dochead{}

\title{Non-perturbative results for large-$N$ gauge theories }
\author{Biagio Lucini}
\address{Physics Department, Swansea University, Singleton Park,
  Swansea SA2 8PP, UK}
\ead{b.lucini@swansea.ac.uk}

\begin{abstract}
 It has been known for a long time that large-$N$ methods can give
 invaluable insights into non-perturbative phenomena such as
 confinement. Lattice techniques can be used to compute quantities at
 large $N$. In this contribution, I review some recent large-N lattice
 results and discuss their implications for our understanding of
 non-perturbative QCD. 
\end{abstract}


\end{frontmatter}


\section{Introduction and motivations}
\label{sect:introduction}
An analytical determination of observables in Quantum Chromodynamics
(QCD) is still an open issue. From the computational point of view, much
progress has been achieved by formulating the theory on a spacetime
lattice and determining physical quantities using Monte Carlo
simulations. Lattice QCD is by now a mature field, which provides a
first-principle framework for computing numerically hadronic
quantities. However, while the results provide a robust evidence (if still needed
at all) that QCD {\em is} the theory describing strong
interactions, unfortunately our ability to compute the spectrum does
not necessarily provides physical insights on the relevant low-energy
phenomena, namely confinement and chiral symmetry
breaking. 

From an analytical perspective, one of the most promising approaches
was provided long ago in~\cite{'tHooft:1973jz}. The key observation is
that if we consider QCD in the general context of SU($N$) gauge theories
and take the limit for the number of colours $N$ going to infinity
keeping constant the 't Hooft coupling $\lambda = g^2 N$ (with $g$
gauge coupling of the SU($N$) theory), the system undergoes a drastic
simplification at the diagrammatic level. In fact, it can be easily
seen that in this limit only the planar diagrams (i.e. the Feynman
diagrams that can be drawn in a plane without crossing lines)
survive. In addition, diagrammatic contributions to observables can be 
arranged in a topological expansion, where the topology of a diagram is
reflected by a well-defined power of 1/$N$ weighting its contribution.  

While from the qualitative point of view the large-$N$ idea allows us to
understand various phenomenological features of QCD, in the strict
quantum field theoretical context it has proven to be still difficult
to arrive at first-principle determinations of observables even in
this simplified framework. Much progress was achieved following the
gauge-string duality conjecture~\cite{Maldacena:1997re}, which
led to the idea of computing non-perturbative quantities in QCD using
the supergravity limit of an appropriate string theory. From the
analytical point of view, this shifted the game to the construction of
a string theory background that is dual to large-$N$
Yang-Mills theory or large-$N$
QCD~\cite{Maldacena:2000yy,Klebanov:2000hb,Sakai:2004cn}. However, the
evaluation of the size of the 1/$N$ corrections is still out of reach
in this framework, since it involves going beyond the supergravity
approximation. Besides, string theory naturally embeds
supersymmetry. This means that, in addition to gluons and quarks, the
gauge theory dual to a string theory will have other fields, whose
effects on the infrared spectrum need to be carefully discussed. 

Although this approach is one of the most popular, the gauge-string
duality is not the only framework that performs analytical
calculations in the large $N$ limit of gauge theories. Among other
frameworks, we mention the topological string model recently proposed 
in~\cite{Bochicchio:2013eda}.

With these premises, a numerical approach to the
large-$N$ limit of SU($N$) gauge theories can serve a twofold purpose:
(a) it provides a first-principle quantification of the deviations of
QCD observables from their large-$N$ limit in the non-perturbative
regime; (b) it can provide a more direct numerical guidance to
calculations aiming at identifying appropriate string theory duals of
QCD. Moreover, analytical progress can be inputed back into
numerical calculations of QCD to inform numerical interpolations or
extrapolations. Inspired by these motivations, following earlier
attempts, in the past fifteen years a broad lattice programme of
numerical simulations has been undertaken with the goal
of providing firm quantitative results for QCD observables using
lattice techniques
(see~\cite{Panero:2012qx,Lucini:2012gg,Lucini:2013qja} for recent 
reviews).  

In this contribution, we review the foundations and the most recent
developments of lattice calculations in the large-$N$ limit of SU($N$)
gauge theories. The rest of the article is organised as follows. In
Sect.~\ref{sect:2} we give a brief overview of the large-$N$ general
results that will be used in our lattice calculations, with the
lattice formulation of the problem exposed
in~Sect.~\ref{sect:3}. Sect.~\ref{sect:4} will be devoted to
the presentation of numerical results for glueballs and mesons. A
brief summary with an overview on future perspectives completes this
work.   

\section{Large $N$ and QCD}
\label{sect:2}
The foundations of large-$N$ gauge theories are the subject of various
pedagogical reviews
(e.g.~\cite{Lucini:2012gg,Lucini:2013qja,Coleman:1980mx,Manohar:1998xv}). Here
we will briefly present an overview of the line of
arguments and of the main results, referring to the literature for a
more in-depth tractation.

In a SU($N$) gauge theory with $N_f$ fundamental fermion flavours, let
us rescale the fields with $N$ in such a way to expose the gauge
coupling $\sqrt{\lambda} = g \sqrt{N}$~\cite{Coleman:1980mx}. This
allows us to easily track the factor of $N$ contributions
in Feynman diagrams. In particular, one finds that (a) a vertex
contributes a factor of $N$; (b) a fermionic loop contributes a factor
of $N$; (c) a fermionic propagator contributes a factor of $1/N$. In
addition, at large $N$, one can consider a gluon line as a double line
with two orientations, corresponding respectively to a fermion and to
an antifermion. In this double-line notation, the above considerations
about colour contributions provided by fermions naturally extend
to gluons. As a result, for a generic connected vacuum amplitude
${\cal A}$ we find 
\begin{eqnarray}
\label{eq:1}
{\cal A} \propto N^{N_V - N_P + N_L} \ ,
\end{eqnarray}
where $N_V$ is the number of interaction vertices, $N_P$ the number of
fermionic propagators (considering each gluon propagator as two
fermion propagators) and $N_L$ the number of fermion loops (again,
considerings the gluonic contributions as due to fermions and antifermions) in
the corresponding Feynman diagram. Drawing the latter diagram in the
double line notation, one notices that it can be seen as a polygon, or
better, as the surface of a three dimensional solid, with the
arrows on the fermionic lines giving orientation to its flat faces. In
this context, the exponent of the power of $N$ in Eq.~(\ref{eq:1}) is
the Euler characteristic $\chi$:
\begin{eqnarray}
\chi = N_V - N_P + N_L = 2 - B - 2H \ ,
\end{eqnarray}
where $B$ (the number of holes) and $H$ (the number of handles) are
topological invariants of the solid associated with the diagram. Hence,
if $B = H = 0$, for instance, we have a polyhedron, which in this
context has the topology of a sphere. Likewise, removing one face will
pinch a hole in the sphere. Hence, a natural
topological classification of vacuum to vacuum connected diagrams emerges. The
diagrams with spherical topology are dominant at large $N$, and correspond to vacuum
to vacuum processes with only gluonic contributions. Each fermion loop
pinches a hole in the sphere, causing a suppression of $1/N$, while
more complicated drawings (corresponding to crossing of lines) can be
associated to handles, each of which suppresses the diagram by $1/N^2$. 

The diagrams for processes involving glueballs and mesons can be
obtained from vacuum diagrams considering the operators that create
those states as coupled to external sources. One can choose the
normalisation so that that two-point functions of glueballs and
two-point functions of mesons are of order one in the large-$N$
limit, and hence correlators of those states are finite as $N \to
\infty$. The argument can be further developed, leading to the
following considerations:
\begin{itemize}
\item in the pure gauge theory, corrections to the large-$N$ limit can
  be expressed as a power expansion in $1/N^2$, while if fermions are
  present the power series is in $1/N$;
\item quark loop effects are of order $1/N$;
\item amplitudes involving three or more glueball or meson operators
  are zero at $N = \infty$ (i.e. scattering and decays are suppressed
  at large $N$); 
\item the mixing between glueballs and mesons is of order $1/\sqrt{N}$;
\item processes with initial and final quark states involving
  annihilation of all initial quarks in intermediate 
  states\footnote{In QCD, the suppression of those processes is known
    as the Okubo-Zweig-Izuka (OZI) rule.} are forbidden at $N = \infty$. 
\end{itemize}
Hence, the drastic diagrammatic simplification of the theory is
reflected by well-defined physical signatures. Large-$N$
arguments can also be extended to baryons ~\cite{Dashen:1993jt}, for
which a rotor-like spectrum naturally emerges~\cite{Jenkins:1993zu}.

A remarkable feature of the large-$N$ limit, which can easily derived
from the counting rules provided above, is that in the large-$N$
limit the effect of fermion loops disappear. The resulting theory
consists of probe external fermions interacting with the gluons, but
not causing any back-reaction on the system. At finite $N$, the
approximation that removes all the fermion loops is called the
quenched approximation (which is hence exact in the large-$N$ limit).
The quenched approximation has played an important role in early
numerical simulations of QCD.  

The large-$N$ behaviour of SU($N$) gauge theories (including quenching
of fermionic matter, observed in lattice simulations of SU(3) for
quark masses at which unquenching effects would have been expected to
show up clearly) is closely reminiscent of the physics of QCD,
and in particular of strong suppressions or long lifetimes with
respect to naive evaluations of strengths of couplings and decay
widths. However, these diagrammatic arguments do not address
crucial  questions such as: (1) Can we define rigorously the large-$N$
limit of SU($N$) gauge theories? (2) If this limit exists, how can we
quantify how close it is to QCD? (3) Are large-$N$ diagrammatic
arguments valid also in the non-perturbative regime of QCD?

Answering those questions requires a first principle approach. In the
following section we will show how Lattice Gauge Theory can provide
the needed {\em ab-initio} framework to address these issues.

\section{Lattice formulation}
\label{sect:3}
The lattice discretisation of an asymptotically free gauge theory
provides a non-perturbative gauge invariant regularisation of that
theory that can be used to compute (e.g. numerically) $n$-point
correlation functions at any value of the coupling. There is a
rigorous prescription for removing the ultraviolet cut-off (which in
this case is the spacing $a$ of the grid) that allows to define a Quantum
Field Theory in continuum Euclidean spacetime. In fact, the lattice
prescription can be used as a constructive definition of the Quantum
Field Theory. A programme based on these ideas has been carried out
for QCD in the past forthy years. The framework has reached a level of
maturity such that first principle precision calculations of QCD
observables now begin to be possible. 

While phenomenology suggests to put a substantial effort in the $N =
3$ case and to specialise tools and techniques for real-world QCD, the
arguments exposed in the previous section provide a robust case for investigating
generic SU($N$) gauge theories. The lattice action of a SU($N$) gauge
theory can be written as
\begin{eqnarray}
S = S_g + S_f \ ,
\end{eqnarray}
where $S_g$ is the contribution of the gauge fields and $S_f$ contains
the fermion contribution. The request for constructing a lattice action is that in
the ultraviolet regime (which, by asymptotic freedom, corresponds to
weak coupling) it flows to the perturbative Gaussian fixed point of the
continuum action. This leaves the freedom to add irrelevant terms,
which takes the form of operators of mass dimension $\Delta$ larger
than four. At tree level, these operators are suppressed as $a^{\Delta
  - 4}$. Asymptotic freedom guarantees
that even when loop corrections are taken into account these operators
do not spoil the correctness of the continuum limit.

In order to preserve gauge invariance on a lattice, we formulate the
gauge fields in terms of parallel transports along links connecting
nearest-neighbour lattice points:
\begin{eqnarray}
U_{\mu}(i) = \mathrm{P} \exp \left( i g_0 \int_i ^{i + \hat{\mu}}
 A_{\mu}(x) \mathrm{d} x^{\mu} \right) \ ,
\end{eqnarray}
where $g_0$ is the bare coupling. $i$ is the set of integer
coordinates labelling the given point on a grid, $A$ is the gauge
field in the continuum and the path ordered exponential is taken along
the link stemming from $i$ and ending in $i + \hat{\mu}$, with
$\hat{\mu}$ versor in direction $\mu$. Local SU($N$) gauge
transformations $G(i)$, which have supports on points $i$, transform
the links as follows:
\begin{eqnarray}
U_{\mu}(i) \to \left(G(i)\right)^{\dag} U_{\mu}(i) G(i + \hat{\mu}) \ .
\end{eqnarray}
The path ordered product of links around the elementary square of the
lattice originating from $i$ in positive directions $\hat{\mu}$ and
$\hat{\nu}$ is given by the plaquette variable
\begin{eqnarray}
U_{\mu \nu}(i) = U_{\mu}(i) U_{\nu}(i + \hat{\mu}) \left( U_{\mu}(i + \hat{\nu})
\right)^\dag \left( U_{\nu}(i) \right)^\dag \ ,
\end{eqnarray}
with the negative links connecting $i$ and $i - \hat{\mu}$ identified
with the dagger of the positive links connecting $i - \hat{\mu}$ and
$i$. The simplest choice for $S_g$ is the Wilson action
\begin{eqnarray}
S_g = \beta \sum_{i, \mu < \nu} \left( 1 - {\cal R}\mathrm{e Tr}U_{\mu
    \nu}(i) \right) \ ,
\end{eqnarray}
which is defined in terms of the real parts of the plaquettes summed
over the whole lattice and is weighted by the lattice coupling $\beta
= 2 N/g_0^2$. 

Concerning the fermionic action $S_f$, a naive discretisation
produces doublers (i.e. 15 unwanted species in the continuum
limit). In fact, it has been shown that no fermion discretisation can
be performed such that chirality, absence of doublers and
ultralocality are preserved at the same
time~\cite{Nielsen:1980rz}. The Wilson formulation, which will be used
in this work, break explicitly chiral symmetry. In general, we can
write the action as a quadratic form in the fermion fields
$\psi_{\alpha}(i)$, where $\alpha$ is a spinor index, as follows:
\begin{eqnarray}
S_f = \overline{\psi}_{\alpha} (i) M_{\alpha \beta}(ij)
\psi_{\beta}(j) \ .
\end{eqnarray}
In the Wilson formulation, the operator $M$ (referred to as the Dirac
operator) is given by
\begin{eqnarray}
\\ \nonumber
  M_{\alpha \beta}(ij) &=&  \left(m+ 4r \right) \delta_{ij}
  \delta_{\alpha \beta} \\ \nonumber
  &-& \frac{1}{2} \left[\left(r - \gamma_{\mu}\right)_{\alpha
      \beta}U_{\mu}(i) \delta_{i,j+\mu}  \right.\\
     &+& \left. \left( r +
      \gamma_{\mu}\right)_{\alpha \beta} U_{\mu}^{\dag}(j)\delta_{i,i-\mu}
  \right] \ ,
\end{eqnarray}
where the explicit chiral symmetry breaking is due to $r \ne 0$. In
our simulations, we set $r = 1$.  The explicit breaking of chiral
symmetry determines a non-zero additive renormalisation for the
fermion mass, which needs to be determined as a part of the Monte
Carlo simulations (e.g. by tuning the mass of the pseudoscalar to
zero). The path integral of the theory reads
\begin{eqnarray}
Z = \int \left( {\cal D} U_{\mu}(i)\right)
    (\det M(U_{\mu}))^{N_f} e^{-S_g(U_{\mu \nu}(i))} \ ,
\end{eqnarray}
where the integration over the fermion fields has been performed and
$N_f$ is the number of fermion flavours. 

Non-perturbative results for observables in the pure gauge sector (for
which, $M = \mathbb{I}$) and in the theory with fermionic matter
can be performed at any $N$ and for values of the lattice spacing $a$
for which the theory is close to its continuum limit. At fixed $N$, a 
controlled $a \to 0$ extrapolation can be performed, giving non-perturbative results
for the continuum theory. Finally, each continuum observable can be
extrapolated to $N \to \infty$. For the latter extrapolation, we use a
power series in $1/N^2$ for the pure gauge theory and theories with
non-backreacting fermionic probe matter and in $1/N$ if there are
fermion loops.  The order of the maximum power that is constrained by
our data will give a quantitative characterisation of how close the
theory is to its $N = \infty$ limit. Note that this could depend on
the observable. 

The limits $a \to 0$ and $N \to \infty$ commute~\cite{'tHooft:2002yn}. Due to
computational demands, it is not always practical to perform first the
limit $a \to 0$ and then the limit $N \to \infty$. In cases where this
procedure is unviable, one can still perform the large $N$ limit at
some fixed value of the lattice spacing $a$, where the common value is
set by fixing the numerical value of a dimensionful 
operator expressed in units of $a$ (generally, $a \sqrt{\sigma}$, with
$\sigma$ the string tension or $a T_c$, with $T_c$ the deconfinement phase
transition). 

Whether the continuum limit is performed before or after the large-$N$
limit, the practicalities of the simulations and the need to keep
finite size discretisation artefacts under control often restrict our maximum $N$
to eight. Hence, the results we will use for the latter extrapolation
will be mostly in the interval $2 \le N \le 8$. A different approach is used
by other authors (see
e.g.~\cite{Narayanan:2004cp,Hietanen:2009tu,GonzalezArroyo:2012fx})
that is based on the idea of reduction (at large $N$ the theory can be
formulated on a single-point
lattice~\cite{Eguchi:1982nm,Bhanot:1982sh,GonzalezArroyo:1982hz})  or
partial reduction (finite size effects disappear at large $N$, hence
the minimal lattice size for which the system is confined is already 
asymptotic for the spectrum~\cite{Narayanan:2003fc}). Using those ideas
allows one to reach larger values of $N$, at the expenses of having different
finite-$N$ corrections (in the case of complete reduction) or less
control on finite-size effects (for partial reduction). These
techniques are complementary to those presented here. Giving a
comparison of the methods and providing a discussion of the
corresponding results is beyond the scope of this work.

\section{The spectrum}
\label{sect:4}

\begin{figure}
  \includegraphics[width=1.1\columnwidth]{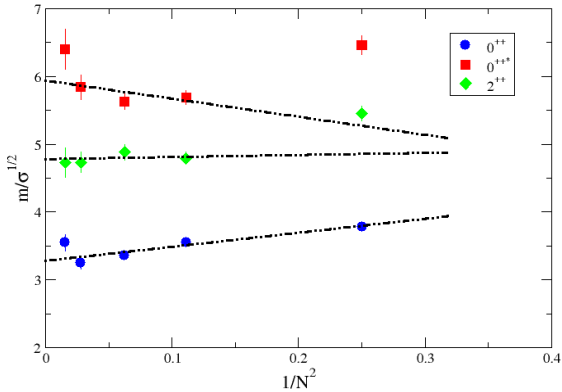}    
\caption{Masses of the lowest-lying glueballs for $N=2,3,4,6,8$,
  with their extrapolation to large $N$~\cite{Lucini:2004my}.}
\label{fig:sungall}
\end{figure}

We start from the computationally easier case of the pure Yang-Mills
theory. Observables of interest in this case include masses of
gauge-invariant states ({\em glueballs}). These are obtained from the
large time behaviour of correlation functions of the form
\begin{eqnarray}
C(t) = \langle O^{\dag}(0) O(t) \rangle \mathop{\propto}^{t \to \infty} e^{- m t} \ ,
\end{eqnarray}
where $O$ is a traced product of links along a closed path ${\cal C}$
that transforms in an irreducible representation of angular momentum
$J$ of the rotational group. For the definition of $C(t)$, the
zero-momentum component (i.e. the spatial average) is taken. If the path is constructed in
such a way that it has a definite parity $P$ and a charge
conjugation eigenstate is constructed by taking either the real
($C = 1$) or imaginary ($C = -1$) part of the trace,
then $m$ is the lowest state in the $J^{PC}$ channel. In practice, after lattice
discretisation, the group of rotations is broken to the dihedral group
of rotations of the cube. Classifying the lattice
states according to the irreducible representations of this group
proves to give a cleaner signal in numerical simulations. The
full rotational quantum numbers can be reconstructed by looking at the
decomposition of the lattice rotational symmetry group under irreducible representations
of $SO(3)$. Likewise, the signal over noise ratio is significantly
improved if in any channel one measures more than one operator and for
each operator its cross-correlations with all others at various $t$.
This defines the correlation matrix
\begin{eqnarray}
C_{lm}(t) = \langle O^{\dag}_l(0) O_m(t) \rangle \ ,
\end{eqnarray}
where $l$ and $m$ label operators associated to two different
paths. The eigenvalues of $C^{-1}(0) C(t)$ decay exponentially, with a rate
controlled by the masses in the given channel. Hence, ordering the
eigenvalues, it is possible to extract the mass of the groundstate and
the mass of the first few excitations, in what can be regarded as a
variational calculation, in which the eigenstates are found by
minimising the hamiltonian over the variational basis provided by the
paths. The number of excitations that one can extract and the accuracy
of the masses crucially depend on the size of the variational basis
and the choice of the operators. The procedure is described in more detail
in~\cite{Lucini:2010nv,Lucini:2014paa}.

\begin{figure}[t]
  \includegraphics[width=0.9\columnwidth]{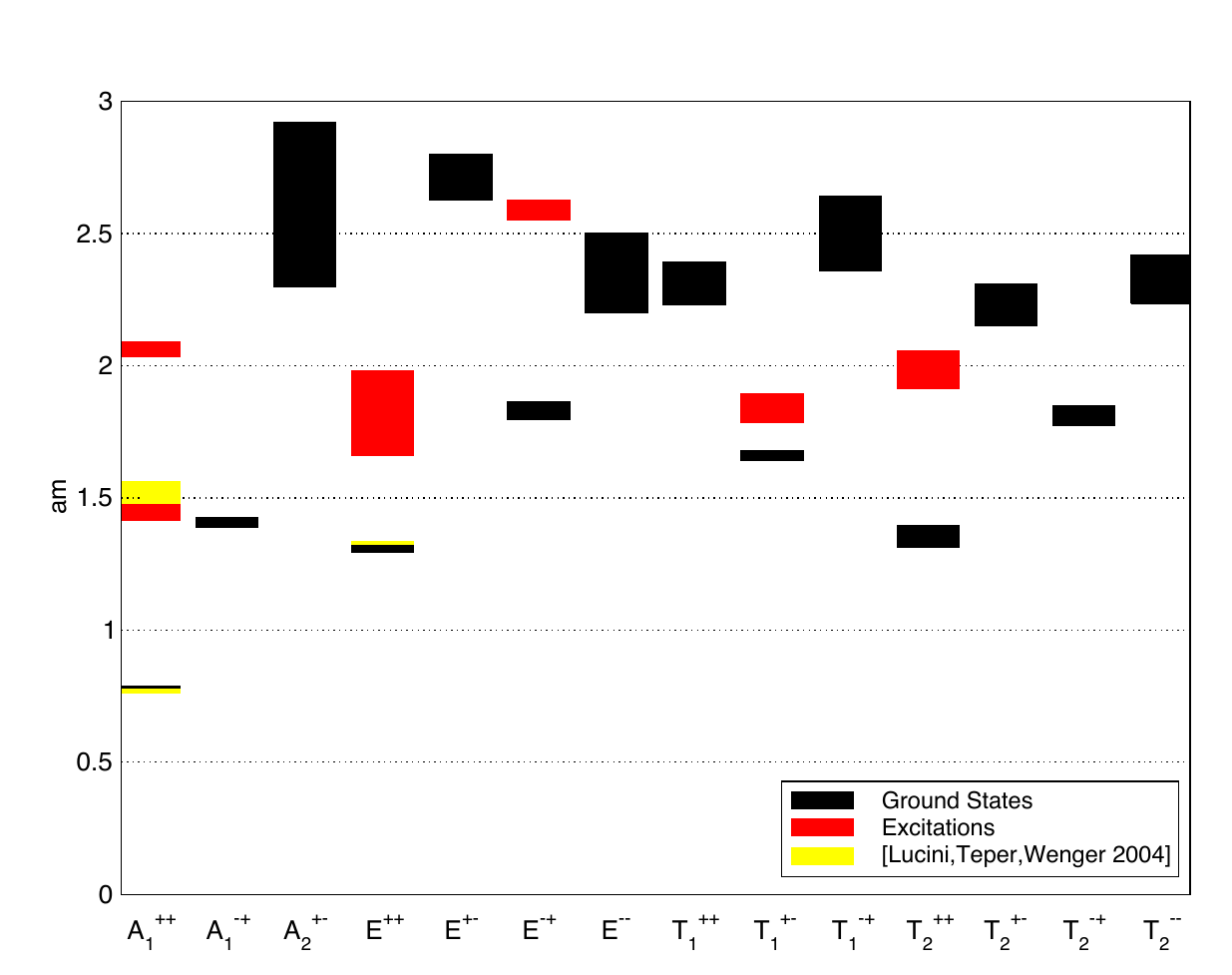}    
\caption{Large-$N$ extrapolation of the spectrum of glueballs at fixed
lattice spacing corresponding to $a T_c = 1/6$ (from~\cite{Lucini:2010nv}).}
\label{fig:glueball_spectrum}
\end{figure}

The continuum spectrum of the lowest-lying glueball was first
extracted in~\cite{Lucini:2001ej}. In Fig.~\ref{fig:sungall} we show the
results of a more recent calculation~\cite{Lucini:2004my}. The results
for the $0^{++}$, its first excitation $0^{++*}$ and the $2^{++}$
glueballs can be summarised by the formulae
\begin{eqnarray}
  \nonumber
  0^{++\phantom{*}}: \qquad \frac{m}{\sqrt{\sigma}} &=& 3.28(8)
  + \frac{2.1(1.1)}{N^2} \ , \\
  0^{++*}: \qquad \frac{m}{\sqrt{\sigma}} &=& 5.93(17) -
  \frac{2.7(2.0)}{N^2} \ , \\
  \nonumber
  2^{++\phantom{*}}: \qquad 
  \frac{m}{\sqrt{\sigma}} &=& 4.78(14) + \frac{0.3(1.7)}{N^2} \ .
\end{eqnarray}
Remarkably, only the leading correction in $1/N^2$ is needed to
describe the data from $N=3$ to $N=8$, with a $\chi^2/\mathrm{dof}$ of
order one, in a fit that gives coefficients of order one. This hints
towards a well-behaved and convergent large-$N$ expansion. This result
provides a quantification of the statement that $N = 3$ is close to $N
= \infty$: a correction $O(1/N^2)$ accounts for the finite value of
$N$ all the way down to $N = 3$ with a level of precision of a few
percents, which is the accuracy of our numerical data.

In order to gain more insight on the glueball spectrum, a more
complete calculation exposing more excitations would be
desirable. In order to eliminate possible sources of systematic
effects, such a calculation should also be able to identify scattering
states and contaminations from finite-size excitations related to loop
wrapping around the periodic lattice ({\em torelons}). Although both
effects disappear in the large-$N$ and large-volume limits, for
a typical calculation their footprint can be non-negligible. A
calculation of this type would require inserting in the variational
basis operators corresponding to the unwanted states, which
unavoidably increases the computational demands and the technical
difficulties. A solution to the latter practical problems has been proposed in~\cite{Lucini:2010nv},
where the construction of the basis operators has been fully
automatised and hence it becomes easier to increase their number. The first
results for the spectrum (this time at fixed lattice spacing) are
reported in Fig.~\ref{fig:glueball_spectrum}. An extrapolation to the
continuum limit is currently being performed.

\begin{figure}[t]
  \includegraphics[width=0.9\columnwidth]{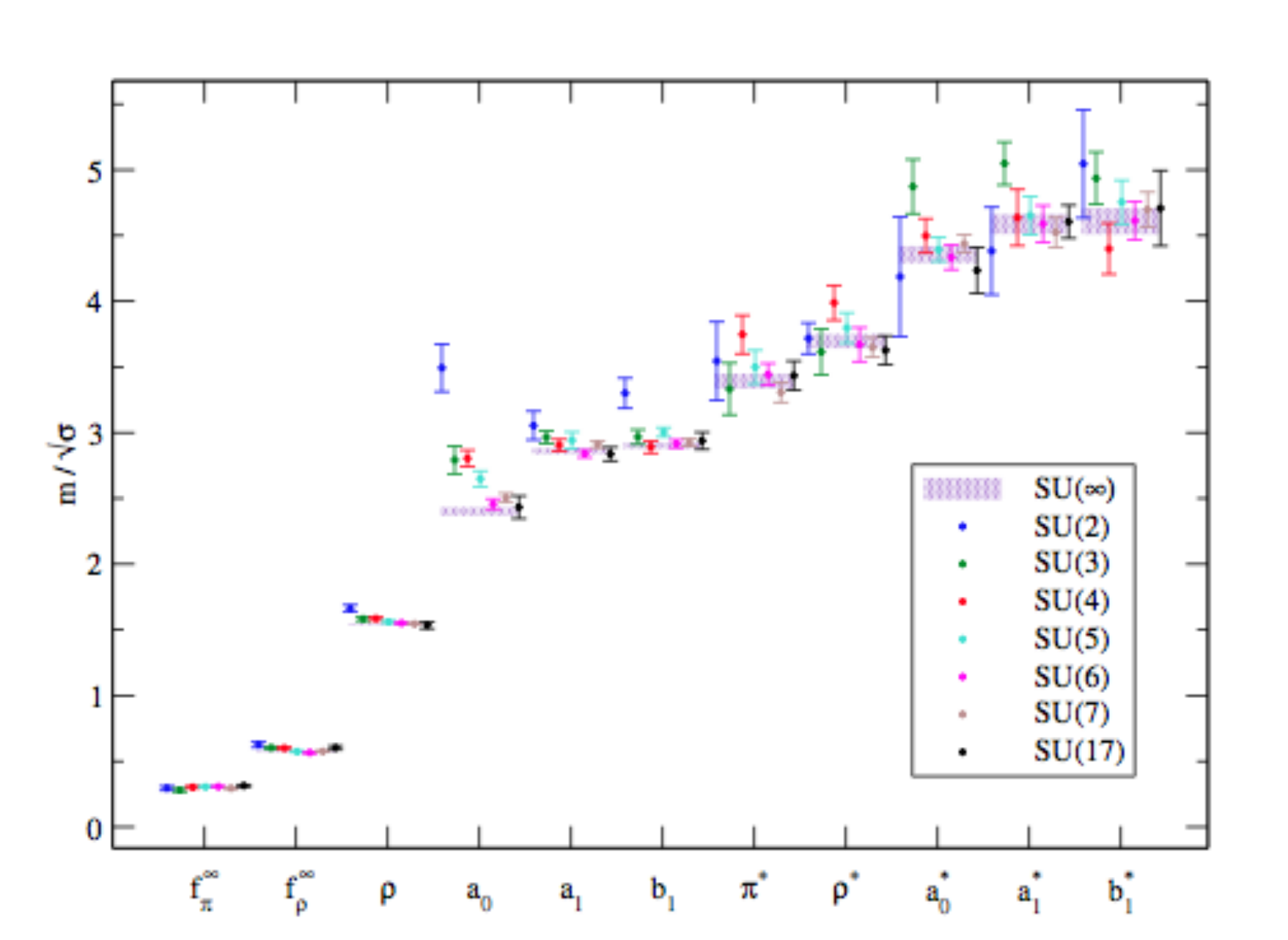}    
\caption{Mesonic observables at various values of $N$ and their
  large-$N$ extrapolation (from~\cite{Bali:2013kia}).}
\label{fig:lnmesonspectrum}
\end{figure}
\begin{figure}[t]
  \includegraphics[width=0.9\columnwidth]{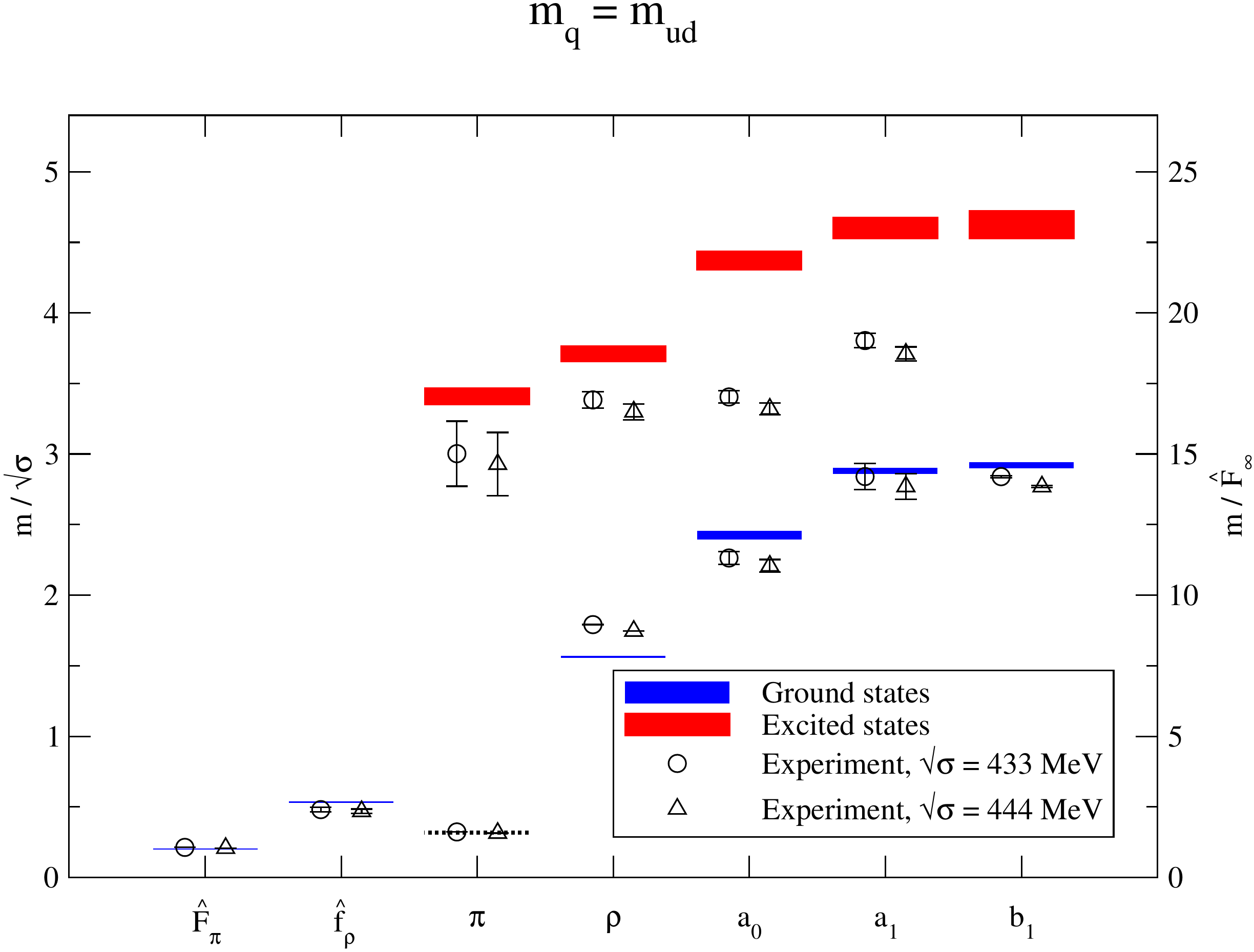}    
\caption{Comparison between lattice large-$N$ results and meson observables
  in QCD (from~\cite{Bali:2013fya}).}
\label{fig:globalExp}
\end{figure}
\begin{figure}[ht]
  \includegraphics[width=0.9\columnwidth]{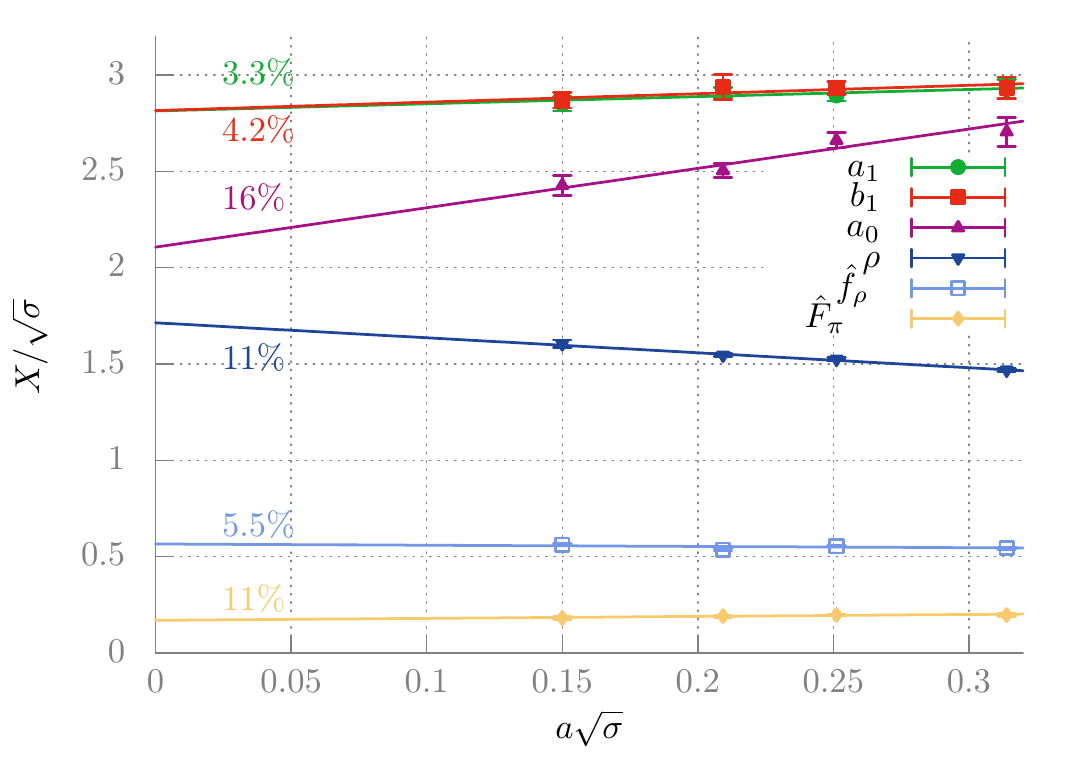}    
\caption{Estimate of lattice size corrections for various meson
  observables in SU(7)  (from~\cite{Bali:2013fya}).}
\label{fig:betafit}
\end{figure}

For fermionic observables, all investigations performed so far are in
the quenched approximation. Since this approximation is exact at $N =
\infty$, it still allows us to obtain the correct values of observables
in that limit. First results for the pseudoscalar and vector
mesons at fixed lattice spacing were reported
in~\cite{DelDebbio:2007wk,Bali:2008an}. These studies were then extended to
decay constants and other mesonic states in~\cite{Bali:2013kia}, with
the continuum limit currently in progress (a status update is reported
in~\cite{Bali:2013fya}). A state of the art determination of the
spectrum at various $N$ showing also the large-$N$ limit is given
in Fig.~\ref{fig:lnmesonspectrum}.

In order to provide a qualitative picture of how close the large-$N$ limit is to real-world QCD, in
Fig.~\ref{fig:globalExp} we show our numerical data together with
experimental data at a quark mass set to its physical value by imposing
that the ratio of the pion mass $m_{\pi}$ and of the pion decay
constant $\hat{F}_{\pi}$ at $N = \infty$ is $m_{\pi}/\hat{F}_{\pi}
= 1.6$, i.e. compatible with the observed SU(3) value. In order to
convert lattice units into MeV, a remaining ambiguity is the value of
the string tension $\sqrt{\sigma}$ (for our calculation, $a
\sqrt{\sigma} = 0.095$). To give a handle on the connected
systematics, two values of $\sqrt{\sigma}$ are reported in
the figure. The lesson one learns is that the deviation of QCD from
its large-$N$ limit is at most 5-7\% for the lowest-lying meson
spectrum and decay constants, while it can be larger (up to 20\%) for
excitations. However, concerning the latter remark, one has to
consider that our lattice calculation has less control of systematic
errors on excited states than it has on groundstates. 

Another potential issue to consider are finite lattice spacing
corrections. First results of a SU(7) study at various lattice
spacings are shown in Fig.~\ref{fig:betafit}. For most of the states,
corrections are of order 5\% and below. However, it is interesting to
note that the $\rho$ and the scalar meson $a_0$ have larger and
opposite corrections that bring their masses close to each other. The
degeneracy of the two states at large $N$ has been argued
in~\cite{Nieves:2009ez}. This example shows that although in most of
the cases we do not expect big surprises arising when taking the
continuum limit, removing lattice spacing corrections in some few
cases can prove to be crucial.

\section{Conclusions and perspectives}
\label{sect:5}
The large-$N$ limit of SU($N$) gauge theories can help us to
understand analytically non-perturbative results in QCD. In order
to make progress with analytical calculations, lattice computations
can be used as a reference. Two examples in different contexts on how
lattice data can be used to inform analytical models are provided
in~\cite{Erdmenger:2007cm,Bochicchio:2013eda} (we refer to those works
for further details). By now, there are various lattice calculations
that provide increasingly solid results in the large-$N$ limit. In
this short review, we have concentrated on the glueball and meson
spectrum. For a wider overview of the field, we refer
to~\cite{Panero:2012qx,Lucini:2012gg,Lucini:2013qja}. 
\section*{Acknowledgements}
I thank C. N{\'u}{\~n}ez for useful comments on the
manuscript and J. Erdmenger and M. Bochicchio for discussions about
their respective results. This work is mostly based on recent original results
obtained in collaboration with G. Bali, L. Castagnini, M. Panero,
A. Rago and E. Rinaldi, whose contributions are gratefully
acknowledged. This research has been partially supported by the STFC
grant ST/G000506/1. The numerical work benefited from computational
resources made available by High Performance Computing Wales and STFC
through the DiRAC2 supercomputing facility.




\bibliographystyle{elsarticle-num}
\bibliography{ichep14bl}

\end{document}